# Observation and analysis of creation, decay, and regeneration of annular soliton clusters in a lossy cubic-quintic optical medium

Albert S. Reyna,[1,*] Henrique T. M. C. M. Baltar,[2] Emeric Bergmann,[3] Anderson M. Amaral,[2] Edilson L. Falcão-Filho,[2] Pierre-François Brevet,[3] Boris A. Malomed,[4,5] and Cid B. de Araújo.[2]

[1]*Programa de Pós-Graduação em Engenharia Física, Unidade Acadêmica do Cabo de Santo Agostinho, Universidade Federal Rural de Pernambuco, Cabo de Santo Agostinho, PE 54518-430, Brazil.*

[2]*Departamento de Física, Universidade Federal de Pernambuco, Recife, PE 50670-901, Brazil.*

[3]*Institut Lumière Matière, UMR5306 Université Lyon 1-CNRS, Université de Lyon, 69622 Villeurbanne Cedex, France.*

[4]*Department of Physical Electronics, School of Electrical Engineering, Faculty of Engineering, and Center for Light-Matter Interaction, Tel Aviv University, Tel Aviv 69978, Israel*

[5]*Instituto de Alta Investigación, Universidad de Tarapacá, Casilla 7D, Arica, Chile*

*Corresponding author. E-mail: areynao@yahoo.com.br

## Abstract

We observe and analyze formation, decay, and subsequent regeneration of ring-shaped clusters of (2+1)-dimensional spatial solitons (filaments) in a medium with the cubic-quintic (focusing-defocusing) self-interaction and strong dissipative nonlinearity. The cluster of filaments, that remains stable over ≈17.5 Rayleigh lengths, is produced by the azimuthal modulational instability from a parent ring-shaped beam with embedded vorticity $l = 1$. In the course of still longer propagation, the stability of the soliton cluster is lost under the action of nonlinear losses. The annular cluster is then spontaneously regenerated due to power transfer from the reservoir provided by the unsplit part of the parent vortex ring. A (secondary) interval of the robust propagation of the regenerated cluster is identified. The experiments use a laser beam (at wavelength 800 nm), built of pulses with temporal duration 150 fs, at the repetition rate of 1 kHz, propagating in a cell filled by liquid carbon disulfide. Numerical calculations, based on a modified nonlinear Schrödinger equation which includes the cubic-quintic refractive terms and nonlinear losses, provide results in close agreement with the experimental findings.

**I. INTRODUCTION**

The stable propagation of self-trapped wave forms, commonly named solitons, has been a topic of great interest in many areas of physics, with fundamentally important realizations in nonlinear (NL) optics, plasmas, and Bose-Einstein condensates (BECs) [1]-[3]. In the spatial domain, permanent shape and size of optical solitons are maintained, in the course of the propagation, by the balance between the linear diffraction and self-interaction in the host medium [1]. The more robust the solitons are, the greater possibilities appear for the design of soliton-based optical devices, chiefly for data transmission and processing [4].

From the theoretical point of view, it is commonly known that a focusing (defocusing) cubic nonlinearity supports stable propagation of one-dimensional [(1+1)D] bright (dark) spatial solitons in nondissipative media [5]. This prediction was corroborated by many experiments that made use of different physical mechanisms to induce the third-order refractive nonlinearity [6]. However, unlike for (1+1)D solitons, stability is a challenge for their two- [(2+1)D] and three-dimensional counterparts, because the critical and supercritical collapse (catastrophic self-focusing) occurs in these settings under the action of the cubic self-focusing, making all the patterns unstable [7].

Some strategies, aiming to avoid the collapse and thus make multidimensional solitons stable, were proposed and supported by experimental results, the most common ones being the use of saturable absorption or high-order nonlinearities for (2+1)D spatial solitons [8], as well as photonic lattices [9]. In particular, robust (2+1)D bright solitons were observed in cubic-quintic [10] and quintic-septimal [11] optical media, where the lower-order nonlinear term is focusing, which is necessary to build the soliton via the balance of the beam's divergence due to linear diffraction, while the higher-order one is defocusing, to secure the arrest of the collapse.

In the same settings, multidimensional solitons with embedded vorticity are subject to the azimuthal modulational instability, leading to spontaneous breakup of the vortex soliton in a set of fragments, the number of which is usually equal to twice the topological charge of the initial vortex [12], [13]. The fragments, during its propagation, move tangentially relative to the initial ring-shaped beam profile, as demanded by the conservation of the orbital angular momentum [14]. Another possible result of the action of the azimuthal modulation instability on the vortex beam is the appearance of a necklace-shaped pattern, built as an annular chain of local intensity maxima, with the thickness of the annulus much smaller than the overall radius [15]. This type of the light structure was also predicted in higher-order TE patterns, which consist of a dark spot surrounded by bright rings, when subjected to asymmetric perturbations, in nonlinear saturable media [16].

Theoretically, it was predicted that optical vortex solitons (OVSs) with all values of the topological charge, $l$, are stable, as solutions to the (2+1)D nonlinear Schrödinger equation (NLSE) with the cubic-quintic nonlinearity, in appropriate intervals of the OVS's propagation constant (the stability intervals are extremely small for high values of $l$) [17], [18]. Furthermore, the same equation gives rise to stable OVS with $l = 1$ in the (3+1)D setting [19]. Experimentally, stable propagation of self-trapped vortex beams over a finite distance was reported in a waveguide with inverted cubic-quintic response, dominated by cubic self-defocusing, acting in the combination with quintic focusing [20].

The search for stability of multidimensional solitons also extends to matter-waves. Recently, remarkable progress has been reported in the creation of multidimensional self-trapped states in BEC. Taking into regard the suppression of the mean-field collapse by quantum fluctuations, stable solitons "quantum droplets" were predicted [21] and experimentally created [22]. Stability of the "droplets" with embedded vorticity was predicted too [23]. Another relevant theoretical result is the prediction of stable "semi-vortices" in two-component spin-orbit-coupled BEC, with vorticity carried by one component [24].

In addition to the quintic and septimal refractive nonlinearities, dissipative NL effects may help to maintain transient stabilization of beams against the collapse [25]. In particular, three-photon absorption (3PA) was fundamental to achieve the stable propagation of (2+1)D bright solitons in liquid carbon dissulfide ($CS_2$) [10] and heavy-metal-oxide glasses [26], as well as to extend a region of intensities where OVSs are effectively stable [13]. However, the dissipative effects cause intensity losses in the course of propagation and, consequently, decay of the spatial solitons, as low-intensity beams cannot develop self-focusing necessary to balance the diffraction.

The present paper reports the observation and analysis of circular chains of filaments originating from optical vortex beams, similar to the above-mentioned necklace patterns, that behave like ring-shaped soliton clusters in two different regions, propagating in a dissipative medium with the cubic-quintic nonlinearity. The two stability regions are identified by considering energy transfer between the filaments and the energy reservoir, provided by the spontaneous azimuthal breaking of the primary vortex beam. Theoretically, the formation and propagation of the soliton cluster is modeled by a modified NLSE, which includes the third- and fifth-order refractive indices, along with the relevant NL extinction coefficient, which characterizes the optical response of liquid $CS_2$ at 800 nm in the regime of femtosecond high-intensity pulsed propagation.

## II. THE EXPERIMENTAL SETUP

The scheme used for the generation and detection of spatial soliton clusters is shown in Fig. 1. The light source used was an amplified titanium sapphire laser with central wavelength 800 nm, which generates ultrashort pulses with 150±5 fs of duration, at a repetition rate of 1.0 kHz. Control of the total power and linear polarization of the incident beam was accomplished by means of a half-wave plate, followed by a Glan-laser polarizer. An optical vortex beam with topological charge $l = 1$ was produced by passing a fundamental TEM$_{00}$ Gaussian beam, previously magnified by a telescope, through a vortex phase plate, manufactured by RPC Photonics. A spatial filter located after the phase plate was used to eliminate higher-order diffracted light. The vortex beam was focused by a lens with focal distance 5 cm (L1) onto the input face of a glass cell filled by liquid CS$_2$.

An imaging system, consisting of a spherical lens (L2) and a CCD camera, which can be translated along the *z*-axis, was used to image the beam profile at the input ($z = 0$) and output ($z = L$) faces of the cell, with magnification $m = 5$. Using this system, it was confirmed that, in all runs of the experiment, the vortex beam at the entrance of the glass cell is structured as a Gaussian background, with full width at half maximum (FWHM) of $w_{0,GB} = 21.7$ μm, containing the vortex core, with FWHM of $w_{0,V} = 6.1$ μm.

## III. THE PROPAGATION OF THE VORTEX BEAM AND FORMATION OF SOLITON CLUSTERS

We examined the propagation of the optical vortex beam in the cell filled with liquid CS$_2$, taking the beam's power below and above the critical value for the onset of the critical collapse in this (2+1)D setting. Figure 2 shows transverse beam profiles at the exit face of the 1.0 cm-long cell.

For low peak intensities ($I_0 \leq 9$ GW/cm$^2$), the vortex-ring (VR) profile at the exit face of the cell exhibits a size 15 times larger than the entrance profile, due to the linear diffraction. Increase of the input intensity makes the nonlinearity significant, which leads to the contraction of the VR's overall radius and thickness, that may be construed as attenuation of the Gaussian background, concomitant with expansion of the vortex core, as shown in Figs. 2(b) and 2(c). Simultaneously, local light intensity grows in the VR.

The VR reaches its minimum thickness for peak intensities around 200 GW/cm$^2$, in which case it is possible to visualize the formation of three small fragments (filaments) in the VR, with diameters approximately equal to the VR thickness. It is known from previous works that the formation of these bright fragments in self-focusing media is a straightforward manifestation of the spontaneous breaking of the azimuthal symmetry induced by the modulational instability [12]. Although the topological charge of the incident vortex beam remains constant ($l = 1$), Fig. 2(d) shows that, at $I_0 = 260$ GW/cm$^2$, the original filaments keep their shape, size and intensity, but the increase of the incident power leads to the formation of new filaments placed along the VR (cf. Ref. [27]), which demonstrates progressive fragmentation of the VR.

The propagation of the VR and formation of filament patterns was further explored in cells with thickness of 2.0 and 5.0 cm, as shown in Figs. 3(a-c) and 3(d-f), respectively. First, Fig. 3(a), corresponding to the input vortex-beam's intensity of 260 GW/cm$^2$, displays filaments with the same size as those observed in the output of the 1.0 cm-long cell [Fig. 2(d)], but with a larger VR radius. This structure remains unchanged for intensities from 350 to 460 GW/cm$^2$, corresponding to Fig. 3(b) and 3(c), respectively. Therefore, despite the fact that the VR keeps expanding in the course of the propagation, we conclude that both the VR's thickness and the diameter of the filaments emerging in the VR keep constant values, for the highest intensities.

On the other hand, Figs. 2(d) and 3(a) show that the number of filaments, for the same intensity of the parent VR, gradually decreases in the course of the propagation. This effect is explained by the energy loss induced by both the multiphoton absorption and scattering of light, which are known features of $CS_2$ at the wavelength of 800 nm [28], as well as by energy transfer between filaments [29] and/or the power reservoir, e.g., the orthogonal VR [30].

Figures 3(d-f) display the transverse beam profile after passing the 5.0 cm long cell. Note that, although the VR's thickness and diameter of the filaments are ~ 2.5 times larger than in Figs. 3(a-c), their dimensions remain constant for intensities from 260 to 350 GW/cm$^2$.

The experimental results obtained for the vortex-beam propagation in cells of thickness 1.0, 2.0 and 5.0 cm are summarized in Fig. 4, across a wide range of peak intensities. Figure 4(a) confirms that, at low intensities, the external (squares) and internal (circles) radii of the VR demonstrate evolution similar to that of the Gaussian beam in the linear limit, whose commonly known form is

$$w_{GB,V}(z) = \left(w_{GB,V}\right)_0 \sqrt{1+\left(z/z_R\right)^2}, \qquad (1)$$

with Rayleigh length $z_R$ and subscript 0 standing for the value at the input face of the cell ($z = 0$). In Fig. 4(a), dependences of $w_{GB}$ and $w_V$, with $z_R \approx 1$ mm, are represented by green lines in the vertical plane. As mentioned above, the nonlinearity of the medium causes gradual contraction (expansion) of the external (internal) radius with the increase of the intensity, until forming a thin VR that maintains its thickness constant for large intensity variations, while the VR's as a whole expands in the course of the propagation. In Fig. 4(b) it is possible to conclude that, after passing the distance of 1 cm, the VR keeps a minimum thickness of 47 μm for peak intensities $I_0 \geq 170$ GW/cm$^2$. Simultaneously, the formation of filaments in the VR was observed at $I_0 > 100$ GW/cm$^2$, reaching a minimum diameter of the filaments, 47 μm, which is equal to the above-mentioned thickness of the VR, also at $I_0 \geq 170$ GW/cm$^2$. In this sense, this region of peak intensities may be identified as a stability range for the formation of the annular multi-filament pattern from the parent VR. Increasing the propagation distance to 2.0 cm, we note that both the VR's thickness and filaments' diameter keep the same values, for the highest intensities ($I_0 \geq 300$ GW/cm$^2$), revealing robust propagation of the established cluster of (2+1)D solitons.

For the cell of length 5.0 cm, the diameter of the (2+1)D solitons that build the cluster increases, but much less than it would grow due to linear diffraction. For this value of the propagation distance, the stability region for the formation of the multi-filament pattern is $I_0 > 250$ GW/cm$^2$, where both the diameter of the filaments (130 μm) and the VR's thickness (150 μm) again stay constant.

## IV. THE MODEL OF THE NONLINEAR PROPAGATION OF THE VORTEX-RING BEAM

The propagation of the vortex beams is adequately modeled by the modified NLSE,

$$2ik\frac{\partial E}{\partial z} + \nabla_\perp E + ik\alpha_0 E = -\mu_0 \omega^2 P^{(NL)}, \tag{2}$$

where $E = E(x,y,z)$ is the envelope of the electric field, $\nabla_\perp = \left(\partial^2/\partial x^2 + \partial^2/\partial y^2\right)$ is the diffraction operator, $z$ is the propagation distance, $k = 2\pi n_0/\lambda = n_0 k_0$, $\lambda$ and $\omega$ are the carrier wavelength and frequency, $n_0$ is the linear refractive index, $\mu_0$ ($\varepsilon_0$) is the vacuum permeability (permittivity), $\alpha_0$ is the linear absorption coefficient, and $c$ is the light speed in vacuum. The term on the right-hand side of Eq. (2) accounts for the contribution of the NL polarization, $P^{(NL)} = \varepsilon_0 \bar{\chi}^{(NL)} E = \varepsilon_0 \left( \chi^{(2N+1)} |E|^{2N} \right) E$, where $\chi^{(2N+1)}$ is the complex $(2N+1)$-th order NL susceptibility, with $N \geq 1$, whose real and imaginary parts represent the refractive and absorptive nonlinearities of the medium, respectively. Then, expressing the optical field in the amplitude-phase form, $E = |E| e^{i\Phi}$, in the simplest approximation, which neglects diffraction, Eq. (2) amounts to evolution equations for the intensity and phase:

$$\frac{\partial \Phi}{\partial z} = \frac{k_0}{2 n_0} \mathrm{Re}\left[ \bar{\chi}^{(NL)} \right], \quad (3)$$

$$\frac{\partial I}{\partial z} = -\frac{k_0}{n_0} \mathrm{Im}\left[ \bar{\chi}^{(NL)} \right] I, \quad (4)$$

with intensity $I = I(R,\theta,z) = 2c\varepsilon_0 n_0 |E|^2$.

Liquid $CS_2$, if excited at 800 nm by femtosecond pulses, behaves as a cubic-quintic (focusing-defocusing) refractive medium, for peak intensities up to hundreds of GW/cm$^2$ [31]. The third- and fifth-order refractive indices, $n_2 = +2.1\times10^{-15}$ cm$^2$/W and $n_4 = -2.0\times10^{-27}$ cm$^4$/W$^2$, which are attributed to the electronic response and effects of molecular collisions, were obtained by means of the well-known Z-scan technique.

As concerns the NL absorption, there are some controversies in the literature regarding its origin. While the first reports on the NL characterization of liquid $CS_2$ associated the intensity dependence of optical losses to the two-photon absorption (2PA) [32], recent works attribute the origin to the three-photon absorption (3PA) [31], [33], which seems to be more consistent with the $CS_2$ energy-level diagram [33]. However, little attention was paid to effects of scattering, which may also cause transmission losses. In this respect, an experimental work that used an integrating sphere to collect the scattered light reveals that the intensity loss may be associated with stimulated scattering, rather than 2PA or 3PA [34].

To elucidate the nature of the intensity losses in liquid $CS_2$, a theoretical wave-coupled model (WCM) was recently proposed to represent the NL response in the case when effects of the NL light scattering are relevant [28], [35]. To validate the WCM in the application to our experimental conditions, we have performed an experiment, measuring the intensity transmitted through a thin sample of liquid $CS_2$ (with thickness 1 mm) versus the input intensity, as shown in Fig. 5. Because diffraction effects are negligible in the thin NL sample, the evolution of the beam's intensity and phase may be modeled by simple equations (3) and (4), in which the nonlinear-refractive part is approximated by the combination of the cubic and quintic terms:

$$\frac{\partial \Phi}{\partial z} = k_0 n_{NL} I = k_0 \left( n_2 + n_4 I \right) I, \tag{5}$$

$$\frac{\partial I}{\partial z} = -\alpha_{NL} I. \tag{6}$$

The experimental data were fitted by solving the differential equation (6), with the intensity-dependent extinction coefficient,

$$\alpha_{NL} = \alpha' \frac{\exp\left((I - I_{th})/I_\delta\right) - \exp\left(-I_{th}/I_\delta\right)}{I/I_{th} + \exp\left((I - I_{th})/I_\delta\right)}, \tag{7}$$

as adopted in the framework of WCM [35]. Values of $\alpha' = 6.4$ cm$^{-1}$, threshold intensity $I_{th}$ =164 GW/cm$^2$, and $I_\delta$ = 22 GW/cm$^2$ were used to provide the best fit for input peak intensities up to 850 GW/cm$^2$ (the red solid line). We stress that WCM not only describes the loss for low intensities, where the 3PA model (with $\alpha_{3PA} = 1.1 \times 10^{-22}$ cm$^3$/W$^2$) may also be used (as shown by the black dashed line in Fig. 5), but it is also capable to model the high-intensity regime, in which the contribution of the scattering leads to highly nonperturbative extinction behavior. Similarly, to [35], the WCM seems to be applicable in the present context by considering the elastic NL scattering due to self-focusing or filamentation.

The comparison of Eqs. (3) and (4) to Eqs. (5) and (6) yields the approximation for the NL susceptibility of liquid $CS_2$ in the form of $\chi^{(NL)} = \left[2n_0(n_2 + n_4 I)I\right] + i\left[n_0 \alpha_{NL}/k_0\right]$. Thus, the modified NLSE, appropriate for modeling the vortex beam propagation, is written as

$$i\frac{\partial E}{\partial z} + \frac{1}{2n_0 k_0}\nabla_\perp E = -\left[k_0(n_2 + n_4 I)I + \frac{i}{2}\alpha'\frac{\exp\left((I-I_{th})/I_\delta\right) - \exp\left(-I_{th}/I_\delta\right)}{I/I_{th} + \exp\left((I-I_{th})/I_\delta\right)}\right]E, \quad (8)$$

where linear losses are ignored because $CS_2$ is transparent at 800 nm.

## V. NUMERICAL SIMULATIONS

The vortex-beam propagation was simulated by numerically solving Eq. (8), using the split-step finite-difference method [36]. Figure 6 shows the resulting evolution of VR profiles for the input peak intensity of 400 GW/cm$^2$, in the course of the passage of the cell filled with $CS_2$, with the total propagation distance of 55 mm. The calculations were initiated with the input waveform $E(R,\theta,z=0) = E_0 \tanh\left[r/(2w_{0,V})\right]\exp\left[-r^2/w_{0,GB}^2 + il\theta\right]$, with power $P = \iint |E(R,\theta,z=0)|^2 rdrd\theta$, where $E_0$ is electric field amplitude, $r = \sqrt{x^2 + y^2}$ and $\theta$ are the polar coordinates in the transverse plane, and $l = 1$ is the topological charge. The intensity profile of the input is displayed in Fig. 6(a). Normally distributed random noise with standard deviation of 3% was added to the initial VR to test the possibility of the onset of the spatial modulational instability, if any, and reproduce the realistic experimental situation.

Figure 6(b) shows strong self-focusing of the VR into a tight shape, concomitant with the emergence of multiple filaments inside the ring, after the passage of the propagation distance equal to 3 mm. From Fig. 6(c) onward, i.e., at $z \geq 6$ mm, the cluster of filaments is completely formed, while the number of filaments gradually decreases in the course of the subsequent long propagation, in agreement with the experimental results presented above for close values of the input intensity and propagation distance, cf. Fig. 3. We stress that, without adding the noise, the simulations produce fission of the vortex beam into just two bright fragments, even for the highest intensities used here, in agreement with previous results for the instability of OVSs with topological charge $l = 1$ [13].

Numerical results were systematically collected for different values of the input intensity. They are summarized in Fig. 7, which displays the VR thickness and filaments' diameter as functions of the propagation distance, for several fixed values of the input power. The variation of the VR thickness with the increase of intensity [Fig. 7(a)] clearly shows the self-focusing effect: the thickness attains its minimum ($\approx 41$ μm) and keeps this value in the interval between $z = 5.5$ mm and $z = 17.2$ mm, for $I = 200$ GW/cm$^2$. Under these same conditions, it is observed in Fig. 7(b) that the filaments keep a constant shape and size in the course of the propagation between $z = 8$ mm and $z = 20.7$ mm. For longer distances, the vortex annulus and filaments carried by it spreads out as a result of the intensity depletion in the presence of the NL loss.

A noteworthy result is found for $I = 400$ GW/cm$^2$, where two intervals of robust self-trapping are observed in Fig. 7, on the scale of the propagation distance. The first interval is 5.5 mm $< z <$ 23 mm, where the VR's thickness and filament diameter keep nearly constant close values, ≈ 41 μm and 37 μm, respectively, The second stability interval is 45 mm $< z <$ 52 mm, in which the thickness and diameter also remain approximately constant, both values being ≈ 177 μm. This second stability interval, which agrees with the experimental findings [see Fig. 4(b)], represents the first evidence of self-regeneration of the annular soliton cluster in dissipative NL media, supported by the power transfer between the bright filaments and background VR.

To further validate this conclusion, we numerically addressed the variation of the integrated power, $P = \iint I(x, y) dx dy$, of the brightest filament and of the entire light field, in the course of the propagation, for the input peak intensity of 400 GW/cm². Figure 8 shows that, in the first interval of the robust propagation, the filament's power slowly decreases, keeping a nearly constant value, around $P_{filament} \approx 0.85$ MW (corresponding to peak intensity ≈ 158 GW/cm² for a filament with the Gaussian transverse profile whose waist is 37 μm), up to $z = 28$ mm. Passing the next 6 mm, strong drop in the filament's power is observed in Fig. 8, causing dramatic increase of the filaments' diameter, as seen in Fig. 7(b). This power loss is caused by the NL extinction, which is significant for values of the filaments' intensity that the simulations demonstrate at this stage of the evolution. Subsequently, at 34 mm $< z <$ 43 mm, sudden increase in the peak power is observed, which is explained by the power transfer from the VR background to the bright filaments, until their power again attains a value close to the above-mentioned value, $P_{filament} \approx 0.85$ MW. The regeneration of the filaments' power signals the entrance into the second interval of the robust propagation of the soliton cluster.

From this analysis, we infer that the critical (threshold) power for the onset of the self-trapping must be slightly smaller than $P_{filament}$, so that the drop of the filament's power to values below $P_\text{filament}$, and its rise back to values above $P_\text{filament}$, which effectively occurs in Fig. 8, explains the sudden expansion of the soliton cluster and its regeneration in the second interval of the robust propagation, which is shown by the red curves in Figs. 7(a) and 7(b). We note that $P_\text{filament}$ is roughly three times greater than the critical power calculated in Ref. [37], $P_\text{cr} = 1.8962\left(\lambda^2/4\pi n_0 n_2\right) \approx 0.28\,\text{MW}$. The difference is reasonable since $P_\text{cr}$ was calculated for a medium with the cubic-only nonlinearity (no quintic term).

## VI. CONCLUSION

We have reported, for the first time to our knowledge, that propagation of light beams in the dissipative NL (nonlinear) medium makes it possible to observe two different intervals of robust propagation of an annular cluster of spatial solitons (filaments), produced by splitting of the parent OV beam, whose residual power is used as a feedback reservoir. As a proof-of-principle, we have implemented the setting experimentally, and developed its numerical simulations, for the robust propagation of a ring-shaped chain of filaments created from a VR (vortex ring) via the development of its azimuthal modulational instability in the optical material ($CS_2$) with the cubic-quintic refractive nonlinearity and extinction dominated by the NL light scattering at high intensities of light. The NL loss in the numerical model is taken as per the wave-coupled model [28], [35]. Experimental and numerical results demonstrate that the annular chain of filaments behaves like a soliton cluster when its power exceeds the critical value necessary for the onset of self-trapping. The robust propagation is maintained by compensation of the NL loss by the power transfer from the reservoir provided by the unsplit part of the parent VR. The presence of the two intervals of the robust propagation of the soliton cluster is explained by the drop of the power below the critical value, followed by its rise back above this value, due to the power supply from the parent vortex beam.

As an extension of this work, it may be interesting to study the influence of the initial vortex beam parameters on the control of the formation, decay and regeneration processes of cluster solitons, as well as to consider the formation and evolution of annular soliton clusters created by the azimuthal modulational instability of VRs with higher vorticities, $l \geq 2$. One promising point of potential studies in this direction may be a possibility of the creation of a robust pattern carrying multiple vorticity, which is an issue of considerable fundamental interest [18]. As concerns possible applications, stable vortical beams (including ones built as circular clusters) with $l \geq 2$ feature a tubular structure with a large inner radius, which can be used as an effective optical conduit steering the transfer of material particles or guiding the propagation of additional quasi-linear probe beams (see, e.g., Ref. [38]).

**ACKNOWLEDGMENTS**

This work was supported by Brazilian agencies Conselho Nacional de Desenvolvimento Cientifico e Tecnológico (CNPq), Fundação de Amparo à Ciência e Tecnologia do Estado de Pernambuco (FACEPE) and Coordenação de Aperfeiçoamento de Pessoal de Nível Superior (CAPES). The work was performed in the framework of the National Institute of Photonics (INCT de Fotônica) project, PRONEX/CNPq/FACEPE and CAPES-COFECUB program. The work of B.A.M. is supported, in part, by the Israel Science Foundation through grant No. 1286/17 and PrInt-CAPES Program (Brazil). This author appreciates hospitality of Departamento de Física at Universidade Federal de Pernambuco (Recife, Brazil).

**Figure captions**

1. (color online) The experimental setup: polarizer (P); telescope (T); vortex phase plate (VPP); mirror (M); spatial filter (SF); spherical lenses (L1 and L2) with focal distance $f_1 = f_2 = 5$ cm.

2. (color online) Transverse beam profiles of the optical vortex, with topological charge $l = 1$, after passing the 1.0-cm-long cell containing liquid $CS_2$, for the following values of the input intensity: (a) 9, (b) 100, (c) 200, and (d) 260 GW/cm$^2$.

3. (color online) Experimentally observed transverse beam's profiles at the output face of the cell with: (a-c) thickness 2.0 cm and intensities (a) 260, (b) 350, and (c) 460 GW/cm$^2$; and (d-f) thickness 5.0 cm, and intensities (d) 260, (e) 290 and (f) 350 GW/cm$^2$.

4. (color online) Intensity dependence of (a) external (squares) and internal (circles) radii of the vortex ring (VR), and (b) its thickness and filaments' (F) diameter for propagation distances 1.0, 2.0 and 5.0 cm (which are represented by dark and light black, red, and blue symbols, respectively). Green lines show the evolution of the widths $w_{GB}$ and $w_V$ of the Gaussian background vortex' core, respectively, as predicted by Eq. (1), with Rayleigh length of ~1 mm, in the linear limit.

5. (color online) Transmittance of $CS_2$ versus the peak input intensity, experimentally measured in the 1-mm-thick cell. The (solid) red and (dashed) black lines indicate the best fits to the data using the WCM and 3PA, respectively (see the text). The inset represents the intensity dependence of the NL extinction coefficient in the framework of the WCM.

6. (color online) The simulated evolution of the vortex-beam profile for the following values of the propagation distance: (a) 0 (the input); (b) 3; (c) 6; (d) 15; (e) 25; (f) 32; (g) 43; and (h) 55 mm. The input intensity is 400 GW/cm$^2$. Note that different scales of the transverse coordinates, ($x,y$), are used in different panels.

7. (color online) Results of numerical simulations showing (a) the vortex-ring's thickness and (b) the diameter of filaments in the annular array as functions of the propagation distance, where $z_R \approx 1$ mm. Dashed vertical lines designate positions for which the experimental results are presented in Figs. 2-4. The numbers next to the curves represent the peak intensity in GW/cm$^2$.

8. (color online) Numerical results for the evolution of the peak power of one filament and the total structure (including the background vortex ring and the cluster of the filaments) for the propagation distance equal to 60 mm, with input intensity 400 GW/cm$^2$. Note the discontinuity of the vertical axis in the figure.

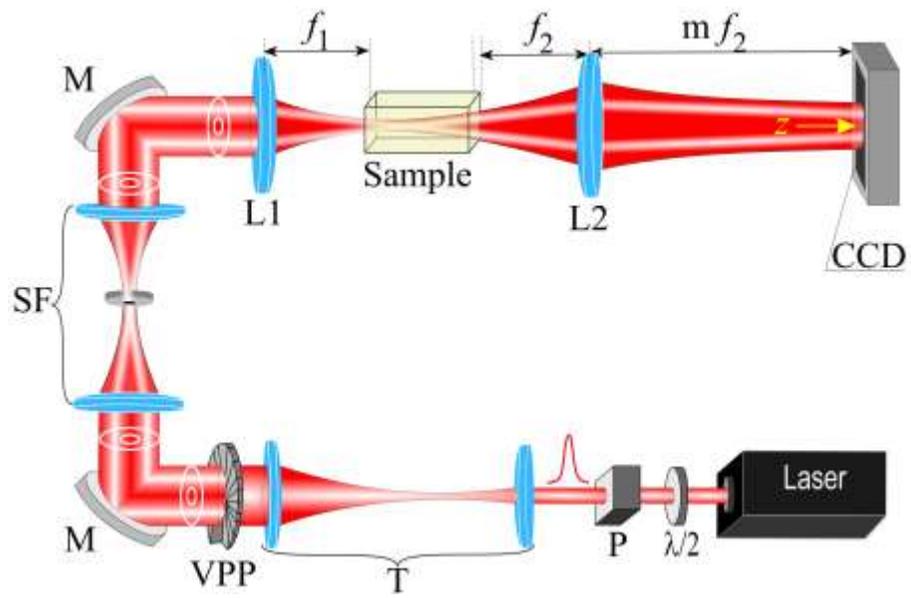

Fig. 1 Reyna et al.

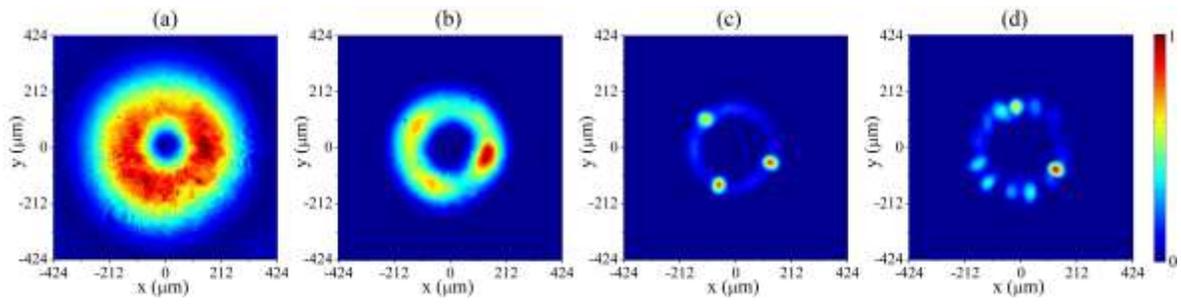

Fig. 2 Reyna et al.

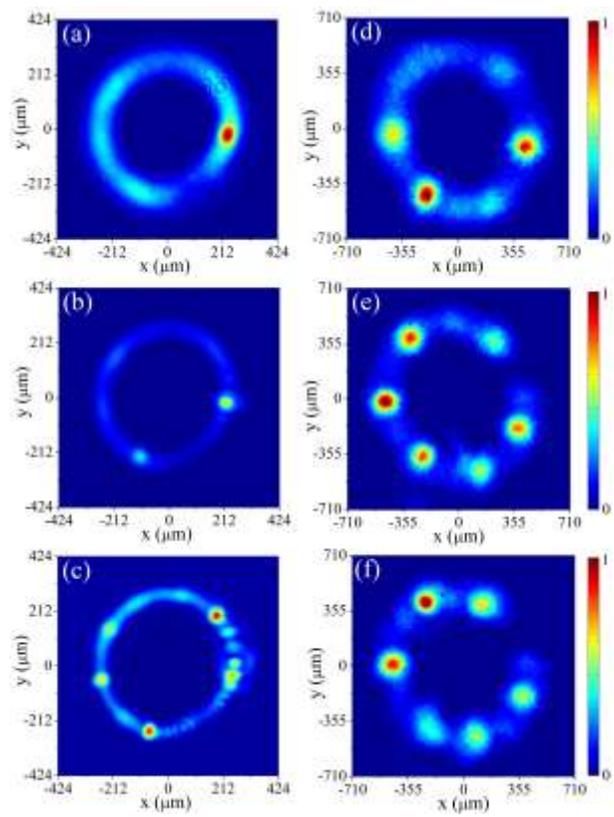

Fig. 3 Reyna et al.

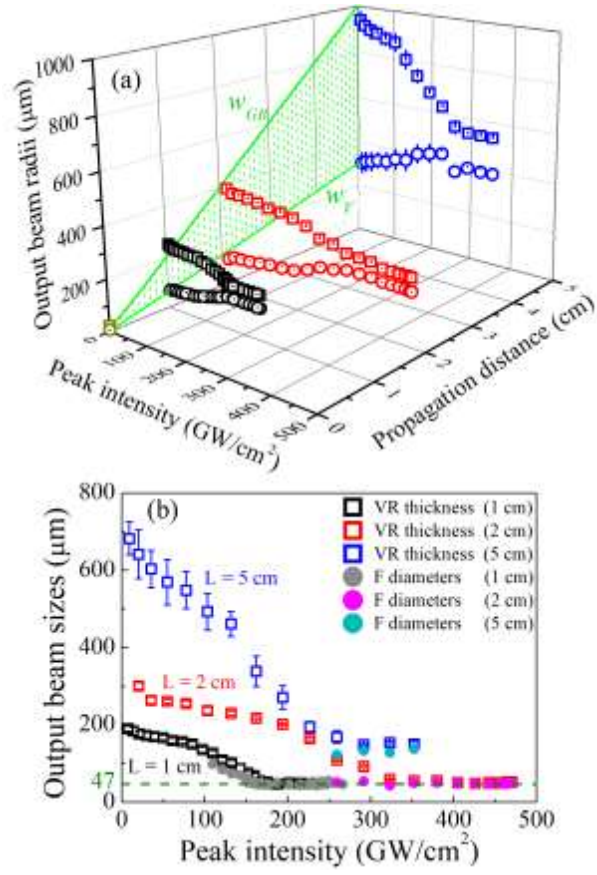

Fig. 4 Reyna et al.

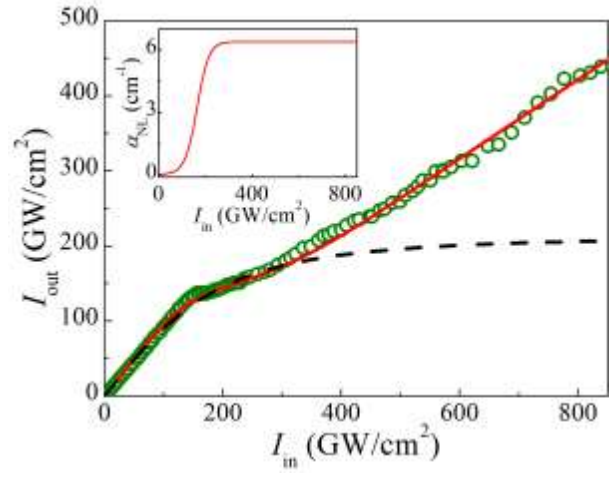

Fig. 5 Reyna et al.

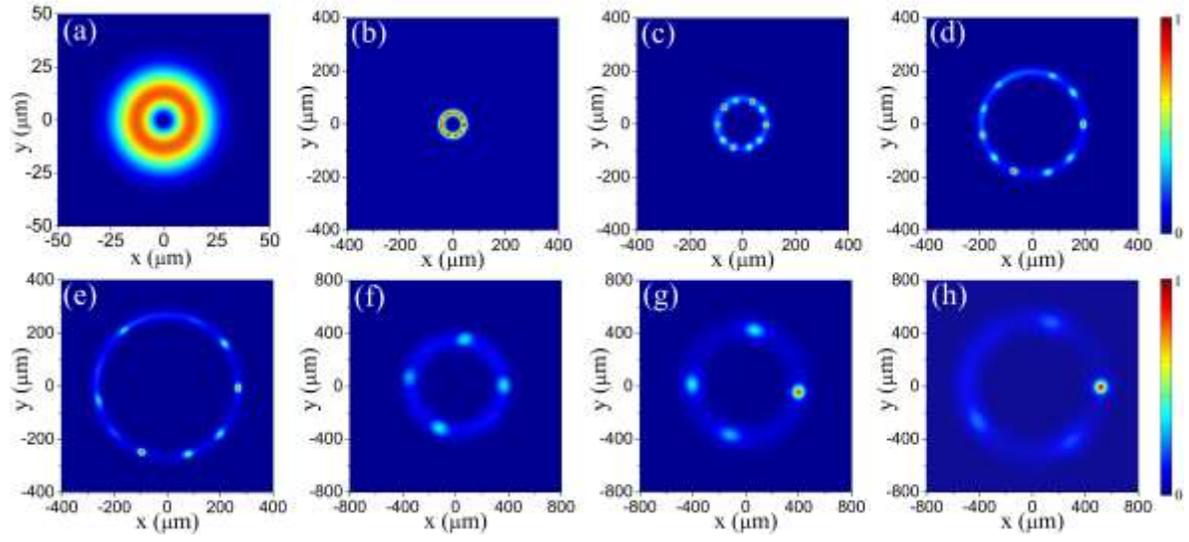

Fig. 6 Reyna et al.

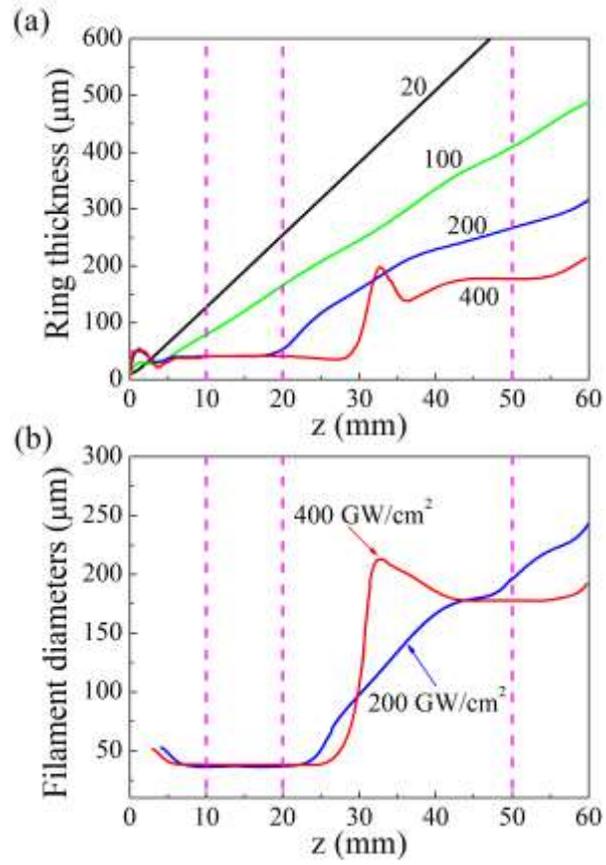

Fig. 7 Reyna et al.

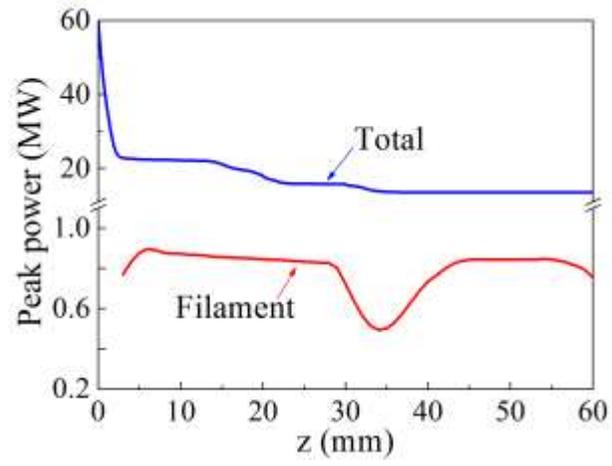

Fig. 8 Reyna et al.